\documentstyle[12pt]{article}
\baselineskip=\normalbaselineskip
\ifx\TwoupWrites\UnDeFiNeD\else\target{\magstepminus1}{11.3in}{8.27in}
	\source{\magstep0}{7.5in}{11.69in}\fi
\newfont{\fourteencp}{cmcsc10 scaled\magstep2}
\newfont{\titlefont}{cmbx10 scaled\magstep2}
\newfont{\authorfont}{cmcsc10 scaled\magstep1}
\newfont{\fourteenmib}{cmmib10 scaled\magstep2}
	\skewchar\fourteenmib='177
\newfont{\elevenmib}{cmmib10 scaled\magstephalf}
	\skewchar\elevenmib='177
\makeatletter
\newcommand\nonsequentialeqnum{
	\@addtoreset{equation}{section}
	\def\theequation{\arabic{section}.\arabic{equation}}}
\newif\ifp@bblock  \p@bblocktrue
\newcommand\nopubblock{\p@bblockfalse}
\newcommand\topspace{\hrule height 0pt depth 0pt \vskip}
\newcommand\p@bblock{\begingroup \tabskip=\hsize minus \hsize
	\baselineskip=1.5\ht\strutbox \topspace-2\baselineskip
	\halign to\hsize{\strut ##\hfil\tabskip=0pt\crcr
	\the\Pubnum\crcr\the\date\crcr}\endgroup}
\newcommand\YITPmark{\hbox{\fourteenmib YITP\hskip0.2cm
        \elevenmib Uji\hskip0.15cm Research\hskip0.15cm Center\hfill}}
\renewcommand\titlepage{\ifx\TwoupWrites\UnDeFiNeD\null\vspace{-1.7cm}\fi
	\YITPmark\vskip0.6cm
	\ifp@bblock\p@bblock \else\hrule height 0pt \relax \fi}
\makeatother
\newtoks\date
\newtoks\Pubnum
\newtoks\pubnum
\Pubnum={YITP/U-\the\pubnum}
\date={\today}
\newcommand{\frontpageskip}{\vspace{12pt plus .5fil minus 2pt}}
\renewcommand{\title}[1]{\frontpageskip
	\begin{center}{\titlefont #1}\end{center}\par}
\renewcommand{\author}[1]{\frontpageskip\par\begin{center}
	{\authorfont #1}\end{center}
	\nobreak
	}

\newcommand{\address}[1]{\par\begin{center}{\sl #1}\end{center}\par}

\renewcommand{\thanks}[1]{\footnote{#1}}
\renewcommand{\abstract}{\par\frontpageskip\centerline{\fourteencp Abstract}
	\vspace{8pt plus 3pt minus 3pt}}
\newcommand\YITP{\address{Uji Research Center \\
	       Yukawa Institute for Theoretical Physics\\
               Kyoto University,~Uji 611,~Japan\\}}
\begin{document}
\thispagestyle{empty}
\newcommand{\bb}{\begin{equation}}
\newcommand{\ee}{\end{equation}}
\renewcommand{\a}{\alpha}
\renewcommand{\pd}{\partial}
\renewcommand{\O}{{\hat O}}
\renewcommand{\L}{\Lambda}
%

%\nopubblock        %% uncomment in making submit-version
%\nonsequentialeqnum %% uncomment in (Section.Number) equation number style.
\pubnum{93-34 \cr hep-th/9311047}
\date{November 1993}
\titlepage

\title{Ward Identities of ${\bf W}_{\infty}$ Symmetry in Liouville Theory
coupled to $c_M < 1$ Matter}

\author{Ken-ji HAMADA\footnote{E-mail
address: hamada@yisun1.yukawa.kyoto-u.ac.jp}}

\YITP

\abstract{We investigate the Ward identities of the ${\rm W}_{\infty}$
symmetry in the Liouville theory  coupled to the $(p,q)$ conformal matter.
The correlation functions are defined by applying the analytic
continuation procedure for the matter sector as well as the Liouville
one. We then find that the Ward identities are equivalent to the
${\rm W}_q$ algebra constraints deduced from the matrix model.
}

\newpage

\indent

  The exact solutions of the matrix models have provided us with a
great deal of information on two dimensional quantum
gravity~\cite{gm,ds,bk}.
One of the prominent features of 2D quantum gravity coupled to
$c_M < 1$ conformal matter  is the appearance of the nonlinear
structures called the W and the Virasoro constraints~\cite{fkna,dvv}.
In order to clarify the structures of the  spectrum and the
factorization, it is necessary to investigate them in terms of the
continuum approach, namely the Liouville thoery [6--18].
This attempt was  done in~\cite{pb,ha} and developed
in our recent work~\cite{hb}.

  In this article we generalize our arguments from the viewpoint of
the ${\rm W}_{\infty}$ symmetry and derive the W and the Virasoro
constraints completely as the Ward identities of the currents.
Applying the $SO(2,{\bf C})$ rotation procedure~\cite{lzb,cs,cdk},
which preserve the structure of the operator product expansion,
we can obtain the ${\rm W}_{\infty}$ currents for $c_M <1$
theory~\cite{cdk} from
those for $c_M =1$~\cite{w,kp,kle}. The Ward identities of ${\rm
W}_{\infty}$ symmetry for $c_M =1$ theory are discussed in~\cite{kle}.
Here we consider the Ward
identities for $c_M =1-6(p-q)^2 /pq$ minimal theory.  The notable
point different from $c_M =1$ theory is how the ${\rm W}_q$ algebra is
realized  from the linear ${\rm W}_{\infty}$ algebra.

  To derive the ${\rm W}_q$ algebra constraints we need to fix the
normalization of the scaling operator and the convention of the matter
momentum (or charge) as well as the Liouville one, which give the
one-to-one correspondence  with the matrix model observables. We also
need to define the correlation function by applying the analytic
continuation procedure~\cite{gl,kit} for the matter sector as well.

  The Liouville theory  coupled to the $(p,q)$ minimal conformal
matter is defined by the action
\begin{eqnarray}
  &&  S_0(p,q)
       =\frac{1}{8\pi}\int d^2 z \sqrt{\hat g}({\hat g}^{\alpha\beta}
            \partial_{\alpha}\phi \partial_{\beta}\phi
                 +2i Q_L {\hat R}\phi )
              \nonumber \\
  && \qquad\qquad~
      +\frac{1}{8\pi}\int d^2 z \sqrt{\hat g}({\hat g}^{\alpha\beta}
            \partial_{\alpha}\varphi \partial_{\beta}\varphi
                +2i Q_M {\hat R}\varphi )
\end{eqnarray}
with the background charges
\bb
     Q_L =-i (p+q)Q ~ , \qquad  Q_M = (p-q)Q ~,
\ee
where $Q=1/\sqrt{2pq}$. The scalar fields $\phi$ and $\varphi$ are the
Liouville and the matter fields respectively.  The scaling operators
of this theory are given by
\bb
      O_j = \int d^2 z V_j (z,{\bar z})
          =\int d^2 z {\rm e}^{\a_j \phi(z,{\bar z})}
                      {\rm e}^{i\beta_j \varphi(z,{\bar z})} ~,
\ee
where the momenta are parametrized as
\bb
     \a_j =(p+q-j)Q ~, \qquad \beta_j =(p-q+j)Q ~.
\ee
These fields are identified with the gravitational primaries and their
descendants:
\bb
        O_{nq+k} =\sigma_n (O_k ) ~,\qquad ( k=1, \cdots ,q-1;
                    \quad n\in {\bf Z}_{\geq 0})~,
\ee
where $n=0$ states are gravitational primaries. Note that the
convention for $\beta_j$ is important when we identify the Liouville
theory with the matrix model. We can then show that the edge states
$O_j ~(j=0~{\rm mod}~q)$ decouple.

   The discrete states of ghost number one $R_{r,s}(z)$ with the
parametrization
by two negative integers $r$, $s$ clasified in~\cite{bmp} exist at
the momenta $\a_{-rp-sq}$ and $\beta_{rp-sq}$. The discrete states
$R_{r,s}~ (r,s \in {\bf Z}_-)$ form the ${\rm W}_{\infty}$ algebra. Here we
normalize the fields such that
\bb
     R_{r,s}(z)R_{r^{\prime},s^{\prime}}(w)
       = \frac{1}{z-w}(r s^{\prime} - r^{\prime} s)
          R_{r+r^{\prime}+1, s+s^{\prime}+1}(w) ~.
\ee
There also exist the discrete states of ghost number zero $B_{r,s}(z)$
at the same momenta as $R_{r,s}$. These states have the ring structure
\bb
   B_{r,s}(z)B_{r^{\prime},s^{\prime}}(w)
       = B_{r+r^{\prime}+1, s+s^{\prime}+1}(w) ~.
\ee

   Combining $R_{r,s}$ and ${\bar B}_{r,s}$ we can construct the symmetry
currents
\bb
    W_{r,s}(z,{\bar z})=R_{r,s}(z){\bar B}_{r,s}({\bar z}) ~,
                \qquad r,s \in {\bf Z}_- ~,
\ee
which satisfy
\bb
     \pd_{{\bar z}} W_{r,s}(z,{\bar z})
            =\{ {\bar Q}_{BRST} ,[{\bar b}_{-1},
                              W_{r,s}(z,{\bar z})] \} ~.
\ee

  Let us define the correlation functions of the Liouville theory.
We consider the interaction theory
\bb
       S=S_0(p,q) +\mu \O_1 -t \O_{p+q} ~,
\ee
where $O_1 $ is the dressed operator of the lowest dimensional matter
field  and $O_{p+q}$ is nothing but the screening charge $S^+
=\int{\rm e}^{i\sqrt{2}\beta_+ \varphi}$,
$\beta_+ =\sqrt{p/q}$.\footnote{
 Note that we do not use another
screening charge  $S^- =\int {\rm e}^{i\sqrt{2}\beta_- \varphi}$,
$\beta_- =-\sqrt{q/p}$ because it is not included in the
definition of the scaling operators (5).
}
%%%%%%%%%%%%%%%%%%%%%%
Here we introduce the normalized operators $\O_j ~(j \in {\bf Z}_+)$
defined by
\bb
      \O_j = \L(j) O_j ~, \qquad
         \L(j)=\frac{\Gamma(j/q)}{\Gamma(-j/q)}~.
\ee
Note that $\L(j)~(j>0)$ vanhishes at $j=0~{\rm mod}~q$. We will see
that the operators $\O_j ~(j>0, j\neq 0~{\rm mod}~q)$ directly
correspond to the matrix model obserbables.  After integrating over
the zero modes of the Liouville and the matter fields the correlation
functions of the scaling operators are expressed as the free field
one:
\bb
   \ll \prod_{j \in S}\O_j \gg_g
      = \Bigl( -\lambda \frac{Q}{\pi} \Bigr)^{-\frac{\chi}{2}}
           \mu^s \frac{\Gamma(-s)}{\a_1} \frac{t^n}{n!}
          < \prod_{j \in S}\O_j ~ (\O_1)^s (\O_{p+q})^n >_g ~,
\ee
where $g $ is genus, $\chi =2-2g$ and
\begin{eqnarray}
    & & s =  \frac{1}{p+q-1}[(p+q)\chi -\sum_{j \in S}(p+q-j) ] ~, \\
    & & n =  \frac{1}{p+q-1}[-\chi +\sum_{j \in S}(1-j) ] ~.
\end{eqnarray}
The $\Gamma$-function comes from the zero mode integral of $\phi$. The
zero mode integral of $\varphi$ gives the Kronecker delta which
guarantees the momentum neutrality of matter sector.
The expression connects between the correlators in the interaction
picture $\ll \cdots \gg_g $
and ones in the free picture $< \cdots >_g $. If $s $ and $n$ are
integers, the correlation functions can be
calculated. However $s$ and $n$ are not integers in general. According
to the argument of~\cite{gl,kit} we define the correlators by
analytic continuations in $s $ and $n$, where $n!$ is defined by
$\Gamma (n+1)$.

 In the following we consider the Ward identities of the currents
\bb
       \int d^2 z \partial_{\bar z}
         \ll W_{-k,-n-k}(z,{\bar z})\prod_{j\in S} \O_j \gg_g =0 ~,
           \quad (k=1, \cdots , q-1;~ n \in {\bf Z}_{\geq 1-k})
\ee
which will be just identified with the ${\rm W}^{(k+1)}_{n}$
constraints. The equations for $k=1$ is
the Virasoro constraints and others are the W constraints.

  We first discuss the Ward identity for the current $W_{-1,-n-1}$
and show that it is expressed as the Virasoro condition for the
exponential of the partition function. The explicit form of the
current $W_{-1,-n-1}=R_{-1,-n-1}{\bar B}_{-1,-n-1}$ is given by
\bb
     R_{-1,-n-1}(z)=-(n+1)! H_-(z)
                     {\rm e}^{\Phi(z)-nqQ\phi(z)+inqQ\varphi(z)}
\ee
for $z$ side and
\bb
     {\bar B}_{-1,-n-1}({\bar z})=
         \bigl( {\bar B}_{-1,-2}({\bar z}) \Bigr)^n
\ee
for ${\bar z}$ side, where $H_-(z) $ and ${\bar B}_{-1,-2}$ are
defined by
\begin{eqnarray}
    && H_-(z)=\oint \frac{du}{2\pi i}{\rm e}^{-\Phi(u+z)} ~,  \\
    && \Phi(z)=(p-q)Q\phi(z) +i(p+q)Q\varphi(z)
\end{eqnarray}
and
\bb
   {\bar B}_{-1,-2}({\bar z})
         = \Bigl[ {\bar c}{\bar b} +\sqrt{\frac{p}{2q}}
             ({\bar \pd}\phi +i{\bar \pd}\varphi) \Bigr]
              {\rm e}^{\a_{p+2q}\phi({\bar z})}
              {\rm e}^{i\beta_{-p+2q}\varphi({\bar z})} ~.
\ee
The operator product expansion (OPE) between the current
and the scaling operator is easily calculated as
\begin{eqnarray}
   &&  W_{-1,-n-1}(z,{\bar z}) O_k (w,{\bar w})
            \nonumber   \\
   &&  \quad = \frac{1}{z-w} \frac{(-1)^n}{q^{2n+1}}
          k^2 (q+k)^2 \cdots ((n-1)q+k)^2 (nq+k)
          O_{nq+k}(w,{\bar w})
            \nonumber      \\
   &&  \quad = \frac{1}{z-w} \frac{k}{q} \L^{-1}(k) \L(nq+k)
          O_{nq+k}(w,{\bar w})  ~,
\end{eqnarray}
where $O_k (z,{\bar z}) = {\bar c}({\bar z})c(z)V_k (z,{\bar z})$.
The $\L$ factors are renormalized into the scaling operators so that
we obtain
\bb
         W_{-1,-n-1}(z,{\bar z}) \O_k (w,{\bar w})
           = \frac{1}{z-w} \frac{k}{q} \O_{nq+k}(w,{\bar w}) ~.
\ee
The derivative $\pd_{{\bar z}}$ picks up the OPE singularity and so we
get
\begin{eqnarray}
  &&  0 = \int d^2 z \pd_{{\bar z}} \ll W_{-1,-n-1}(z,{\bar z})
                \prod_{j\in S} \O_j \gg_g
                     \nonumber  \\
  && \quad  = \pi \frac{p+q}{q}
             t \ll \O_{nq+p+q} \prod_{j\in S} \O_j \gg_g
             -  \frac{\pi}{q}
             \mu \ll \O_{nq+1} \prod_{j\in S} \O_j \gg_g
                                \\
  && \qquad  + \frac{\pi}{q}
             \sum_{k\in S} k
                \ll \O_{nq+k} \prod_{j\in S \atop (j\neq k)} \O_j \gg_g
              + \int d^2 z \ll \pd_{{\bar z}} W_{-1,-n-1}(z,{\bar z})
                \prod_{j\in S} \O_j \gg_g  ~.
                    \nonumber
\end{eqnarray}
The first and the second correlators of r.h.s. come from the OPE with
the potentials $\O_{p+q}$ and  $\O_1 $ respectively. Usually the last
correlator would vanish because the
divergence of the current is the BRST trivial (9). However, in this
case,  the boundary of moduli space pinching 2D surface become
dangerous and the last correlator gives  nonvanishing contributions.

  Using the relation of factorization~\cite{hb,s,pa} we can calculate the
contribution from such a boundary,
\begin{eqnarray}
  && -\lambda \frac{Q}{\pi} \sum^{\infty}_{N=0} \sum^{\infty}_{k=1}
      \int^{\infty}_{-\infty} \frac{dh}{2\pi}
         \ll F_1 ~ \int_{|z| \leq 1} d^2 z
            \pd_{{\bar z}} W_{-1,-n-1}(z,{\bar z})
           \nonumber  \\
  && \qquad\qquad
         \times D |-h, \beta_{-k}, N \gg \ll h, \beta_k, N | F_2 \gg ~,
\end{eqnarray}
where $D$ is the propergator. $F_1$ and $F_2$ are sets of operators
composed of ones in $S$.
The integer $N$ stands for the oscillation level of the states.
The zero level state is defined by
\bb
     |h, \beta_k > = {\rm e}^{i(h+Q_L)\phi(0)}
                     {\rm e}^{i\beta_k \varphi(0)}
                      |0>_{L,M} \otimes {\bar c}_1 c_1 |0>_G  ~,
\ee
which is the eigenstate of the Hamiltonian $H=L_0 +{\bar L}_0 $ with
the eigenvalue $h^2 +E_{k,N}$, where
$E_{k,N}= k^2 Q^2 +2N$.\footnote{
Note that, as discussed in~\cite{ct,s,pa}, the spectrum of the
free theory $|h,\beta_k >$ is the same as that of interaction theory
$|h,\beta_k \gg$.}
%%%%%%%%%%%%%%%%%%%
It is normalized as
\bb
    \ll h^{\prime},\beta_{k^{\prime}},N^{\prime}
       | h,\beta_k ,N \gg_{g=0}
          = \frac{-\pi}{\lambda Q} 2\pi\delta(h+h^{\prime})
             \delta_{k+k^{\prime},0}
             \delta_{N,N^{\prime}} ~.
\ee
The zero mode integral of the Liouville field now produces the
$\delta$-function. The propergator is defined by
\bb
      D=\int_{|z|\leq 1}\frac{d^2 z}{|z|^2}
                z^{L_0}{\bar z}^{{\bar L}_0}
       =2\pi \biggl( \frac{1}{H}-\lim_{\tau \rightarrow \infty}
          \frac{1}{H}{\rm e}^{-\tau H} \biggr) ~.
\ee
The last term stands for the boundary of moduli space pinching 2D
surface.  Since the BRST charge commutes with the
Hamiltonian, there is no contribution from $1/H $ term in the
propagator. While the boundary term gives  nonvanishing quantities
 in the limit $\tau \rightarrow \infty$,
\begin{eqnarray}
  &&   \lim_{\tau \rightarrow \infty} \lambda \frac{Q}{\pi}
         \sum^{\infty}_{N=0}
         \sum^{\infty}_{k=1} \int^{\infty}_{-\infty} \frac{dh}{2\pi}
           \int_{{\rm e}^{-\tau}\leq |z| \leq 1} d^2 z
          \ll F_1 ~ [{\bar b}_{-1}, W_{-1,-n-1}(z,{\bar z})]
              \nonumber  \\
  && \qquad \times  {\bar Q}_{BRST}
         \frac{2\pi}{H}{\rm e}^{-\tau H}
             |-h, \beta_{-k},N \gg \ll h, \beta_k,N | F_2 \gg ~.
\end{eqnarray}
Changing the variable to $z={\rm e}^{-\tau x +i\theta} $, where
$0 < x < 1 $ and $0 < \theta < 2\pi $, the integrals of $x$ and the
momentum $h$ can be evaluated exactly in the limit $\tau \rightarrow
\infty$ by using the saddle point method. Here we omit the details of
the calculation (see~\cite{hb}). The result is
\bb
   \lambda \frac{\pi}{q} \sum^{nq-1}_{k=1}
         \L(k) \L(nq-k) \ll F_1 ~ O_{nq-k} \gg \ll O_k ~ F_2 \gg ~,
\ee
where $N \neq 0$ modes vanish exponentially as ${\rm e}^{-2N\tau}$.
The sum of $k$ is restricted within $0 < k < nq$ because the saddle
point of the $x$ integral is now located at $x=k/nq$ which needs to be
within the interval $0<x<1$ to give a nonvanishing quantity.
Furthermore, note that the expression vanishes at $k=0~{\rm mod}~q$
because of the factors $\L(k)$ and $\L(nq-k)$. This factors are
absorbed in the scaling operators and we obtain
\begin{eqnarray}
   &&        \int d^2 z \ll \pd_{{\bar z}} W_{-1,-n-1}(z,{\bar z})
             \prod_{j\in S} \O_j \gg_g
                \nonumber  \\
   &&  = \frac{1}{2!}\lambda \frac{\pi}{q}
             \sum^{nq-1}_{k=1 \atop (k \neq 0~{\rm mod}~q)}  \biggl[
          \ll \O_{nq-k} \O_k \prod_{j\in S} \O_j \gg_{g-1} \biggr. \\
   && \qquad \biggl. + \sum_{S=X\cup Y \atop g=g_1 +g_2}
                  \ll \O_{nq-k} \prod_{j\in X} \O_j \gg_{g_1}
                  \ll \O_k \prod_{j\in Y} \O_j \gg_{g_2} \biggr] ~.
                \nonumber
\end{eqnarray}
The first term of r.h.s. is a variant of the boundary (29), where
a handle is pinched. The factor $1/2!$ corrects for double counting.
Therefore we finally get the equation
\begin{eqnarray}
  && 0=\frac{p+q}{q}t \ll \O_{nq+p+q} \prod_{j\in S} \O_j \gg_g
       +\frac{x}{q} \ll \O_{nq+1} \prod_{j\in S} \O_j \gg_g
        \nonumber   \\
  && \qquad + \sum_{k\in S} \frac{k}{q} \ll  \O_{nq+k}
              \prod_{j\in S \atop (j\neq k)} \O_j \gg_g      \\
  && \qquad +\frac{1}{2}\frac{\lambda}{q}
              \sum^{nq-1}_{k=1 \atop (k \neq 0~{\rm mod}^q)}
             \biggl[  \ll \O_{nq-k} \O_k \prod_{j\in S} \O_j \gg_{g-1}
                 \biggr.   \nonumber  \\
  && \qquad\qquad\qquad \biggr.
                  + \sum_{S=X\cup Y \atop g=g_1 +g_2}
                  \ll \O_{nq-k} \prod_{j\in X} \O_j \gg_{g_1}
                  \ll \O_k \prod_{j\in Y} \O_j \gg_{g_2} \biggr]
                \nonumber  ~,
\end{eqnarray}
where we set $x=-\mu $. This equation is nothing but the Virasoro
constraint~\cite{fkna,dvv}
\bb
      {\rm L}_n \tau \vert_{x_1 =x \atop {x_{p+q}=t
               \atop x_j =0~(j\neq 1,p+q) }}=0~,
        \qquad \tau={\rm e}^{Z(x_1, x_2, \cdots)} ~,
\ee
where
\bb
    {\rm L}_n = \sum_{-k+m=nq} \frac{k}{q} x_k \pd_m
           +\frac{1}{2}\frac{\lambda}{q} \sum_{k+l=nq} \pd_k \pd_l
\ee
and $Z(x_1, x_2, \cdots)$ is the partition function of the Liouville
theory defined by the action $S=S_0 (p,q) -\sum_j x_j
\O_j $. Here $x_j$ and $\pd_j,~ (j=0~{\rm mod}~q)$ are discarded.

  Next we consider the Ward identity of the current $W_{-2,-n-2}$
which will gives the ${\rm W}^{(3)}_n$ constraint~\cite{fkna,dvv}
\bb
     {\rm W}^{(3)}_n \tau \vert_{x_1 =x \atop {x_{p+q}=t
               \atop x_j =0~(j\neq 1,p+q) }}=0~,
\ee
where $\tau$ function is defined in (32) and
\bb
      {\rm W}^{(3)}_n=\sum_{-l-k+m=nq}
                 \frac{l}{q} \frac{k}{q} x_l x_k \pd_m
                   +\frac{\lambda}{q} \sum_{-l+k+m=nq}
                      \frac{l}{q} x_l \pd_k \pd_m
    +\frac{1}{3}\frac{\lambda^2}{q^2}\sum_{l+k+m=nq}\pd_l \pd_k \pd_m ~.
\ee
To derive the single derivative term we need to calculate the OPE
\bb
   W_{-2,-n-2}(z,{\bar z}) O_k (0,0) \int d^2 w V_l (w,{\bar w})
      =\frac{1}{z} C^{(3)}_n(k,l) O_{k+l+nq}(0,0) ~.
\ee
The coefficient for $n=-1$ has already been calculated in~\cite{hb}.
We now normalize the current $W_{-2,-1}$ as
\bb
    W_{-2,-1}= \Bigl[ \pd^2 \Phi + (\pd \Phi)^2 \Bigr]
              \biggl[ {\bar c}{\bar b}+\sqrt{\frac{q}{2p}}
                      ({\bar \pd}\phi -i{\bar \pd}\varphi) \biggr]
             {\rm e}^{\a_{2p+q}\phi}{\rm e}^{i\beta_{-2p+q}\varphi}~.
\ee
We then get
\bb
      C^{(3)}_{-1}(k,l)=2\pi \frac{k}{q} \frac{l}{q}
                        \L^{-1}(k) \L^{-1}(l) \L(k+l-q) ~.
\ee
The coefficient for general $n$ is given by using the ${\rm
W}_{\infty}$ algebra. Noting that
\bb
      W_{-2,-n-2}(z,{\bar z})= \frac{-1}{2n+3}
            [Q_{-1,-n-2}, W_{-2,-1}] ~,  \qquad
         Q_{r,s} =\oint \frac{d z}{2\pi i} W_{r,s}(z,{\bar z})
\ee
and using the results (21) and (38) we easily obtain
\bb
      C^{(3)}_n(k,l)=2\pi \frac{k}{q} \frac{l}{q}
                        \L^{-1}(k) \L^{-1}(l) \L(k+l+nq) ~.
\ee

  The two derivative term in (35) is derived by calculating the
following boundary
\begin{eqnarray}
    && \lim_{\tau \rightarrow \infty}
          -\lambda \frac{Q}{\pi} \sum^{\infty}_{k=1}
            \int^{\infty}_{-\infty}\frac{dh}{2\pi}
         \ll F^{\prime}_1 ~ \Bigl\{
              \int_{{\rm e}^{-\tau}\leq |z| \leq 1}d^2 z
              {\bar \pd}W_{-2,-n-2}(z) \int_{|w| \leq |z|}d^2 w V_l (w)
            \nonumber  \\
   &&\qquad\qquad\qquad
           +\int_{{\rm e}^{-\tau}\leq |z| \leq 1}d^2 z V_l (z)
              \int_{|w| \leq |z|}d^2 w {\bar \pd}W_{-2,-n-2}(w)
                 \Bigr\}      \\
    && \qquad\qquad\qquad \times
            \frac{-2\pi}{H}{\rm e}^{-\tau H}
              |-h, \beta_{-k}\gg \ll h, \beta_k | F^{\prime}_2 \gg ~,
                 \nonumber
\end{eqnarray}
where the primes on $F_1$ and $F_2$ stand for the exclusion of the operator
$O_l$. The integrals of $h$ and $z$ are also evaluated by using the saddle
point method. Using the ${\rm W}_{\infty}$ algebra (39) and the result
for $n=-1$~\cite{hb} we obtain for general $n$
\bb
   2\pi^2 \frac{\lambda}{q} \sum^{nq+l-1}_{k=1}
           \frac{l}{q}  \L^{-1}(l) \L(k) \L(nq+l-k)
            \ll F^{\prime}_1 ~ O_{nq+l-k} \gg
              \ll O_k ~ F^{\prime}_2 \gg ~.
\ee
The three derivative term can be calculated as a variant of the
boundary (42). Prepare another factorization relation (see (24)). As
discussed in our previous work~\cite{hb}, noting that the $1/H$ part
of the propagator gives the on-shell states after
integrating over the intermediate momentum, the contribution
corresponding to the three derivative term can be obtained by replacing
$V_l $ in the expression (41) with $-\lambda (Q/\pi)(1/h_l)V_{-l} \ll
O_l ~F_3 \gg$, where
$1/h_l =\int dh ~(h^2 +E_{l,0})^{-1} =\pi/lQ$,
\bb
     2\pi^2 \frac{\lambda^2}{q^2} \sum_{k+l+m=nq}
          \L(m) \L(k) \L(l) \ll F_1 ~O_m \gg
          \ll O_k ~F_2 \gg \ll O_l ~ F_3 \gg ~.
\ee
Combining the boundaries (36), (42) and (43) and their variants
and taking into account the factor $1/2!$ for (42) and $1/3!$ for (43)
to avoid the overcounting, we finally get the ${\rm W}^{(3)}_n$
constraint.

  In general cases it is necessary to calculate the following operator
product
\bb
     W_{-k,-n-k}(z,{\bar z})O_{l_1}(0,0) \int V_{l_2} \cdots \int V_{l_k}
       = \frac{1}{z} C^{(k+1)}_n(l_1,\cdots,l_k)
                  O_{nq+l_1 +\cdots +l_k}(0,0) ~,
\ee
where the OPE coefficient is calculated by applying the
${\rm W}_{\infty}$ algebra recursively,\footnote{
This coefficient is the same as that calculated for $c_M =1$
theory~\cite{kle} up to the sign factor if we take $k_j =l_j /2q$,
where $k_j$ corresponds to the momentum of the tachyon $T^+_{k_j}$.
}
%%%%%%%%%%%%%%%%%%%%%%%%
\bb
      C^{(k+1)}_n =\pi^{k-1} k! \L(nq+l_1+\cdots+l_k)
                    \prod^k_{j=1}\frac{l_j}{q}\L^{-1}(l_j) ~.
\ee
This OPE corresponds to the single derivative term of ${\rm W}^{(k+1)}_n$
constraint
\bb
      {\rm W}^{(k+1)}_n =\sum_{-l_1 -\cdots -l_k +m=nq}
         \frac{l_1}{q} \cdots \frac{l_k}{q}
            x_{l_1} \cdots x_{l_k} \pd_m ~+~\cdots ~.
\ee
We can also calculate the boundary corresponding to the two derivative
term. The terms with more derivatives are caluculated as variants of
the two derivative term.

  We showed that the Ward identities of the currents $W_{-k,-n-k}$ are
equivalent to the ${\rm W}^{(k+1)}_n $ constraints which form the
${\rm W}_q $ algebra. Here note that there are no
${\rm W}^{(k+1)}_{-k}$ constraints in the Liouville theory approach.
The Ward identities of $k \geq q$ will become
redundant. The similar argument appears in the matrix model
approach~\cite{fknb,im}.

  We finally comment on the universality class of the models. Until
now we consider the $S_0 (p,q)$ model perturbed by the operators
$\O_1 (p,q)$ and $\O_{p+q}(p,q)$, where $(p,q)$ stands for the $(p,q)$
minimal theory. The same model can be obtained from the
$(p^{\prime},q)$ model defined by the action
\bb
       S=S_0 (p^{\prime},q)+\mu \O_1 (p^{\prime},q)
                           -t \O_{p+q}(p^{\prime},q) ~,
\ee
where $p^{\prime} \neq p$. The special case $p^{\prime}=q+1$ was
discussed in~\cite{hb}.  $\O_1 (p^{\prime},q)$ and
$\O_{p+q}(p^{\prime},q)$ is the first and the $(p+q)$-th scaling
operators of the $(p^{\prime},q)$ theory.
Note that $\O_{p+q}(p^{\prime},q)$ is
no longer the screening charge of the matter sector.
Really this model satisfies the same recursion
relations as those of the model discussed in this paper. It is easily
seen by noting that
all OPE coefficients of the boundary calculations (21), (29), (40),
(42), (43) and (45) are independent of $p$. It is also explained from
the viewpoint of the $SO(2,{\bf C})$ rotation~\cite{lzb,cs,cdk}. These
two models including the potential terms can interchange each other by the
rotation.

\vspace{0.5cm}
\noindent{\bf Acknowlegement:}
  The author wishes to thank S. Mizoguchi for informing me of ref.~\cite{cdk}.
This work is supported by the Yukawa Memorial Foundation.

\end{document}